\documentclass[epj]{svjour}
\usepackage[utf8]{inputenc}
\usepackage[english]{babel}
\usepackage{amsmath,amssymb}
\usepackage{textcomp,gensymb}
\usepackage{graphicx}
\usepackage{cite}

\begin{document}

\title{Phase diagram of the Ising square lattice with competing interactions}

\author{A.~Kalz,~A.~Honecker,~S.~Fuchs \and T.~Pruschke
}                     
%
%
\institute{Institut f\"ur Theoretische Physik,
         Georg-August-Universit\"at G\"ottingen,
         37077 G\"ottingen, Germany}
\date{Received: May 7, 2008 / Revised version: July 18, 2008} 
%
\abstract{
We restudy the phase diagram of the 2D-Ising model 
with competing interactions $J_1$ on nearest neighbour and $J_2$ 
on next-nearest neighbour bonds via Monte-Carlo simulations. 
We present the finite temperature phase diagram and introduce computational 
methods which allow us to calculate transition temperatures close to the critical
point at $J_2 = J_1/2$. 
Further on we investigate the character of the different phase boundaries
and find that the transition is weakly first order for
moderate $J_2 > J_1/2$.
} 
\maketitle
\section{Introduction \label{intro}}
The search for exotic groundstates in two-dimensional frustrated
quantum spin systems is a topic of intense research (see, e.g.,
Refs.~\cite{P:misguich05,P:richter04} for recent reviews).
In this context, the antiferromagnetic $J_1$-$J_2$ spin-1/2 Heisenberg model
on the square lattice has been intensively studied during the past
two decades. Nevertheless, the nature of an intermediate non-magnetic
phase around $J_2 \approx J_1/2$ has remained under debate until
recently (see Refs.~\cite{P:misguich05,P:mambrini06} and references therein).
Among the possible approaches to the antiferromagnetic
$J_1$-$J_2$ Heisenberg model,
quantum Monte-Carlo (QMC) simulations suffer from a severe sign problem
in the region $J_2 \approx J_1/2$. Other approaches include perturbation
theory around the Ising limit \cite{P:oitmaa96,P:singh03}.
A closely related model is given by hard-core bosons on the square lattice
with nearest neighbour (NN) and next-nearest neighbour (NNN) hopping
and repulsion terms \cite{P:batrouni00,P:batrouni01,P:wessel08,P:chen08}.
In this case, there is no sign problem such that QMC simulations
are possible in principle \cite{P:batrouni00,P:batrouni01,P:wessel08,P:chen08}.
However, simple QMC algorithms suffer freezing problems in the intermediate
regime at $J_2 = J_1/2$ which can be traced to a groundstate degeneracy 
of the Ising limit. This motivated us to perform a model study by solving
the related freezing problems in Monte-Carlo (MC) simulations of the Ising
model.

Investigations of the two-dimensional $J_1$-$J_2$ Ising model on the square lattice
have an even longer history than of the corresponding Heisenberg model,
including in particular MC simulations (see, e.g.,
\cite{P:SK79,P:landau80,P:landaubinder80,P:landaubinder85,P:bloete87,P:MKT06,B:landaubinder}).
Nevertheless, certain issues have remained controversial also in
the case of the Ising model, in particular the nature of the finite-temperature phase 
transition for $J_2 > J_1/2$: MC simulations
\cite{P:SK79,P:landaubinder85,P:MKT06} 
and an investigation of the Fisher zeros of the partition function \cite{P:monroe07}
have suggested non-universal critical exponents, whereas a variational
approach \cite{P:lopez93} and a differential operator technique \cite{P:AVS08}
predict a first-order transition for $J_1/2 < J_2 \lesssim J_1$. In this paper
we resolve this issue in favour of a weak first-order transition
at least for not too large $J_2 > J_1/2$ by providing substantially
improved MC results for the phase diagram.

This paper is organized as follows: in section \ref{sec:model} we introduce
the $J_1$-$J_2$ Ising model on the square lattice and discuss its $T=0$
groundstates. The MC simulation methods are described in
section \ref{sec:methods} and results are presented in section \ref{sec:results}.
We conclude with a summary and outlook in section \ref{sec:discussion}.

\subsection{Model \label{sec:model}}
We study the classical Ising model with competing antiferromagnetic
interactions $J_1$ on the NN bonds and $J_2$ on the NNN bonds ($J_i > 0$):
\begin{eqnarray}
H = J_1 \sum_{\text{NN}} S_i S_j + J_2 \sum_{\text{NNN}} S_i S_j \,  ,
\quad S_i =\pm1 \, .
\label{eq:En}
\end{eqnarray}
We will study square lattices of linear extent $L$ with periodic
boundary conditions.

\begin{figure} 
\begin{center}
\resizebox{\columnwidth}{!}{%
  \includegraphics[]{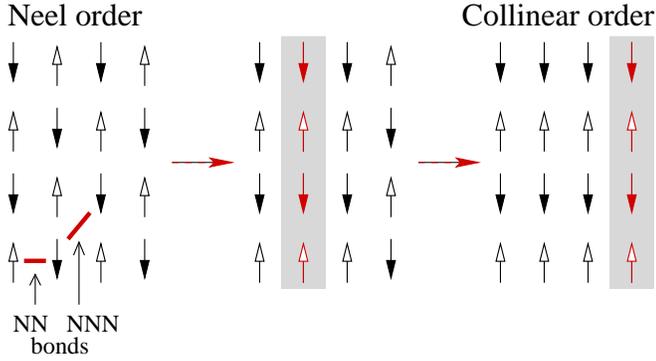}
}
\caption{Sketched are both ordered phases and
a third groundstate configuration at $J_2 = J_1/2$ (middle).
The total degeneracy of the groundstate at $J_2 = J_1/2$ is of order
$2^{L+1}$. Shaded areas mark flipped lines.}
\label{fig:degeneracy}
\end{center}
\end{figure}

For $J_2 = 0$ the groundstate of (\ref{eq:En})
is the known antiferromagnetic solution
which is a N\'eel ordered lattice (see Fig.~\ref{fig:degeneracy}, left)
with energy $E = -2 J_1 N$ ($N = L^2$ is the number of sites).
Switching on the repulsive interaction on the diagonal (NNN) bonds
of the square lattice yields an increase
of the groundstate energy for the N\'eel-ordered state:
\begin{eqnarray}
E_{\text{N\'eel}} = -2 N (J_1-J_2) \, .
\end{eqnarray}
For large $J_2$ the system orders in the collinear (or superantiferromagnetic)
phase (see Fig.~\ref{fig:degeneracy}, right) where all diagonal bonds
are antiferromagnetic, while half of the NN bonds is antiferromagnetic
and the other half is ferromagnetic. Thus, the energy in this state depends
only on $J_2$:
\begin{eqnarray}
E_{\text{Coll}} = -2 N J_2 \, .
\end{eqnarray}
The critical point separating these two phases lies at $J_2 = J_1/2$, where
the transition temperature is suppressed to $T=0$. At this point the
groundstate is highly degenerate. More precisely, there are 
$2^{L+1}-2$ groundstates which can be obtained as follows: flipping a line
of antiparallel spins costs no energy at $J_2 = J_1/2$. Thus, one
can generate almost $2^{L+1}$ groundstates from a N\'eel state by flipping
either the $L$ horizontal or the $L$ vertical lines independently
(the second N\'eel state can be reached in both ways).
In particular, one can reach a collinear state through a series
of intermediate groundstates by $L/2$ line flips, as sketched in
Fig.~\ref{fig:degeneracy}.
Accordingly, close to the critical point
$J_2 = J_1/2$ the energy landscape is characterized by many
local minima which are separated by large
energy barriers. Therefore, MC
simulations using only single-spin flips
have problems to reach the configuration with global minimum energy. 
To solve this problem and to obtain transition temperatures also in 
the vicinity of $J_2 = J_1/2$ we implemented improved MC algorithms 
which we will discuss in the next section.

\section{Methods \label{sec:methods}}
\subsection{Computational methods \label{sec:comp}}
We first tried  to simulate the model with competing interactions
using conventional single-spin flip MC simulations \cite{B:landaubinder}.
However, near the critical point this algorithm does not provide proper results and suffers severe freezing problems in the proximity of the critical temperature and below. To overcome these problems we implemented a parallel tempering algorithm \cite{P:huku96,P:marinari98,P:hans97,P:trebst06}:
a number of simulations with the same set of parameters ($J_1$, $J_2$, $L$) but varying temperatures are simulated simultaneously. After a sufficiently large number of sweeps over the complete lattice an additional MC step proposes a configuration swap between neighbouring simulations with an acceptance rate $p(i,i+1)$:
\begin{eqnarray}
p(i,i+1) = \min\left\{1, e^{\Delta \beta \Delta E}\right\} \, ,\\
\Delta \beta = \frac{1}{T_{i+1}}-\frac{1}{T_i} \, ,\quad
\Delta E = E_{i+1}-E_i  \, .\nonumber
\end{eqnarray}
We have chosen the temperatures $T_i$ logarithmically with a density maximum near the estimated 
transition temperature. A selfadjusting temperature set 
\cite{P:trebst06} was not necessary for our purposes.
To manage the additional computational workload of the parallel tempering
algorithm we implemented a parallelisation of the MC code
via OpenMP \cite{B:omp00,B:omp07}. To assure the independence of the simulations 
we used the \textsc{sprng} (version 4.0) \cite{P:masca00} lagged Fibonacci generator for producing
independent random number streams for each simulation.
We have checked in some samples that our MC results do not change
if we replace the random number generator
by the Mersenne Twister algorithm \cite{P:matsu98}, as implemented in
the \textsc{boost} libraries version 1.33.1.

Another way to improve the MC simulations is to allow not only single-spin
updates but also flipping whole lines of spins in an additional MC step (as
sketched in Fig.~\ref{fig:degeneracy}). This method simulates directly the transition
between degenerate groundstates at $J_2 = J_1/2$ and helps to minimize statistical 
errors for lattice sizes up to $50 \times 50$. For larger lattices the probability to flip a whole line decreases rapidly for ratios $J_2 \neq J_1/2$ while the simulation time to calculate them increases with $L$.
We also tried cluster updates \cite{P:wolff89} but in the case of competing
interactions this method does not help: when one approaches the critical
temperature, the clusters extend over the whole lattice and therefore do
not support the ordering process.

The parallel tempering algorithm gives rise to correlations between different temperature points 
in a simulation which can affect the independency of the calculated data. 
Therefore, the meanvalues and errorbars of the shown observables are derived 
from at least 10 independent MC runs.

\subsection{Order parameters \label{phys}}
\begin{figure} 
\begin{center}
\resizebox{\columnwidth}{!}{%
  \includegraphics[]{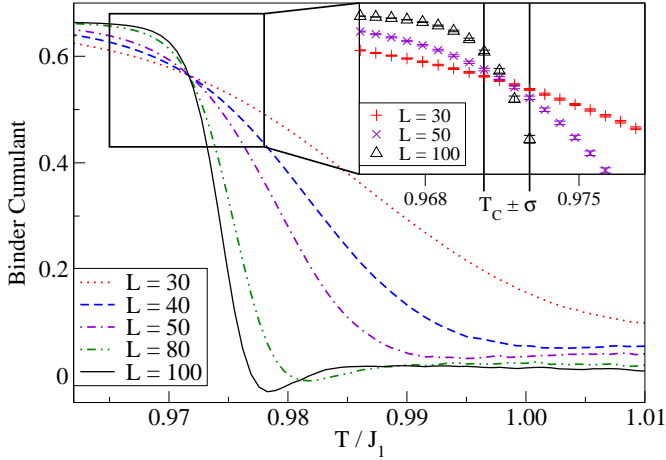}
}
\caption{Binder cumulants depending on temperature for $J_2 = 0.6 J_1$ and  different lattice sizes. In the inset three cumulants are plotted with errorbars. The intersection area gives $T_C$ with small error.}
\label{fig:cumulants} 
\end{center}
\end{figure}

To distinguish the two ordered and the disordered phases we use the respective structure factors
\begin{eqnarray}
S(\vec q)=\frac{1}{N}\sum_{i,j}e^{i\vec q\cdot(\vec x_i-\vec x_j)}\langle
S_i  S_j \rangle
\end{eqnarray}
where $\vec q$ is a vector in momentum space and indicates the different magnetic phases:
\begin{eqnarray}
\vec q = (0,0) &\rightarrow& \text{ferromagnetic order,} \nonumber \\
\vec q = (0,\pi), (\pi,0) &\rightarrow& \text{collinear order,} \\
\vec q = (\pi,\pi) &\rightarrow& \text{N\'eel order.} \nonumber 
\end{eqnarray} 
We identify $M(\vec q) = \sqrt{S(\vec q)/N}$ as our order parameter.
The N\'eel order parameter can be calculated as staggered magnetization due to a simple sublattice rotation.
Using the order of the collinear phase we can also calculate the associated order parameter
via the column or line index of the underlying lattice.
To find the transition temperature we calculate the fourth order
Binder cumulant \cite{P:binder81a,P:binder81b,B:landaubinder} 
for different system sizes $L$:
\begin{eqnarray}
U_4 = 1 -\frac{\langle M^4\rangle}{3\langle M^2\rangle^2}\, .
\end{eqnarray}
For large enough $L$ they will meet in a single point at $T_C$
(for an example see Fig.~\ref{fig:cumulants}).
To analyze the character of the phase transition we calculated the specific heat
and recorded time series of the energies to set up histograms
\cite{P:challa86,P:berg91,P:borgs92}.
Furthermore we calculated the temperature
derivative of the Binder cumulant which is related with the critical exponent
$\nu$ \cite{B:landaubinder} to study the critical behaviour of the system:
\begin{eqnarray}
a L^{1/\nu} = \left.\frac{\partial U_4}{\partial T}\right|_{T=T_C}\,.
 \label{e:derivU4} 
\end{eqnarray}
To have a closer look on the critical point $J_2 = J_1/2$ we
analyzed the peaks of the specific heat via polynomial fitting.

\section{Results}
\label{sec:results}
\begin{figure} 
\begin{center}
\resizebox{\columnwidth}{!}{%
  \includegraphics[]{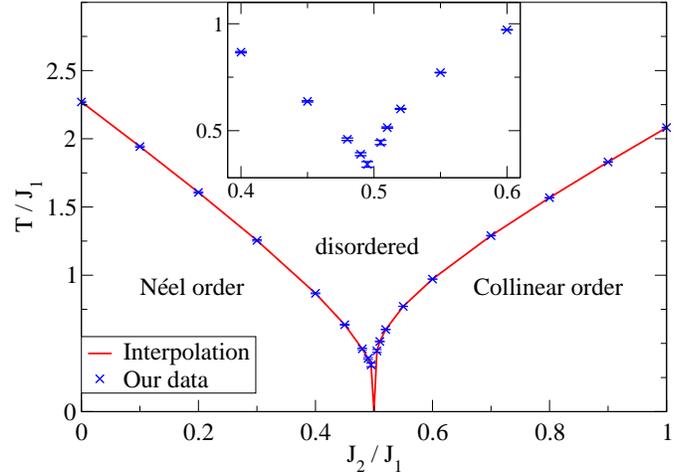}
}
\caption{Plotted are $T_C$'s over the ratio $J_2/J_1$. Data were produced by \textit{parallel tempering MC  simulations}. Near the critical point $J_2 = J_1/2$ additional \textit{line flip updates} led to a better
 convergence.}
\label{fig:phase}
\end{center}
\end{figure}

We calculated the critical temperatures for various ratios of $J_2 /J_1$ especially close to the critical point (Fig.~\ref{fig:phase}). Our data is in good agreement 
with MC results from Landau and Binder for $J_2 \geq 0.6J_1$ \cite{P:landaubinder85}
and $J_2 \leq 0.4J_1$ \cite{P:landau80}.
In addition, our improved MC algorithm with the above described parallel tempering mode
and line flip updates enabled us to obtain results much closer to
$J_2 = J_1/2$ ($|J_2/J_1 - 0.5|= 0.005$) for system sizes up to $N = 250~000$.

\begin{figure} 
\begin{center}
\resizebox{\columnwidth}{!}{%
  \includegraphics[]{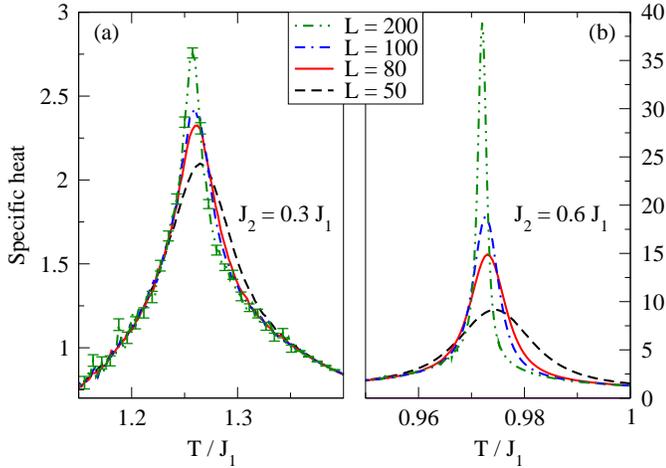}
}
\caption{Comparison of specific heats for $J_2 < J_1/2$ (a) and $J_2 > J_1/2$ (b). The peaks emerge at the critical temperature obtained from the fourth order cumulant calculations.
Note the different magnitudes of the divergence.
Errorbars are omitted if they are of the same order as the linewidth.}
\label{fig:heat0306}
\end{center}
\end{figure}

To investigate the character of the phase transition we calculated
the specific heat and energy histograms for different lattice sizes.
Looking at the peaks in the specific heat for different $J_2/J_1$
suggests different character of the phase transition.
For $J_2 < J_1/2$ (Fig.~\ref{fig:heat0306}(a)) we have a
slowly emerging peak with growing system size.
Indeed, for an Ising like transition 
we expect only a logarithmic divergence in the specific heat\cite{P:onsager44}
and a critical exponent $\nu = 1$ \cite{P:fisher67}. To verify the
Ising like behaviour we had a closer look at the derivative of the Binder
cumulant $U_4$ near the critical temperature. We expect a linear
correspondence in $L$ for equation (\ref{e:derivU4}).
An analysis of our results at $J_2 = 0.3J_1$ yields $\nu = 0.99(1)$
which is in good agreement with the expected value $\nu = 1$.
On the other hand for $J_2 > J_1/2$ and in particular for the
case $J_2 = 0.6 J_1$ shown in Fig.~\ref{fig:heat0306}(b)
there are different opinions about the character of the phase transition.
Some authors assume a continuous phase transition with non universal
exponents \cite{P:SK79,P:landaubinder85,P:MKT06,P:monroe07}
whereas other authors find a first order phase
transition \cite{P:lopez93,P:AVS08}. Note that our results for the specific
heat (Fig.~\ref{fig:heat0306}(b))
differ quantitatively from the approximate results of
\cite{P:lopez93}. 
In particular, the maximum of the specific heat continues to diverge for growing $L$. We estimated the area under  the peak in dependence on the lattice size and find it to converge towards a constant value for large enough $L$. This indicates a $\delta$-peak structure for the specific heat in the thermodynamic limit as expected for a first order phase transition. We should nevertheless mention that the value of the specific heat does not follow the finite size scaling law expected for first order transitions \cite{P:challa86} very well, which may be due to crossover phenomena. 

\begin{figure} 
\begin{center}
\resizebox{\columnwidth}{!}{%
  \includegraphics[]{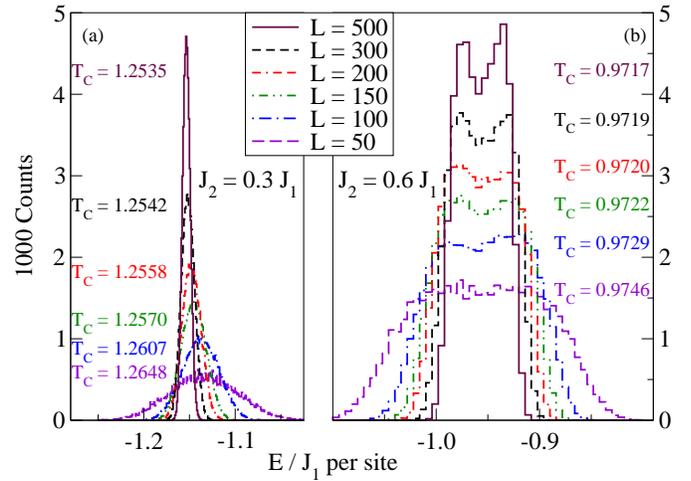}
}
\caption{Energy histogram for $J_2 = 0.3J_1$ ($bin size = 10^{-4} J_1$) and $J_2 = 0.6J_1$ ($bin size= 
6\cdot10^{-3} J_1$). For each lattice size $L$ the histogram is plotted at the critical temperature $T_C(L)$. 
The slowly emerging double peak structure in panel (b)
in comparison to the clearly single peaked structure in panel (a)
indicates a weak first order transition for $J_2 > J_1/2$.}
\label{fig:histo}
\end{center}
\end{figure}

As further verification of our characterisation of the order of the phase transition we computed energy histograms. 
First, we extracted a system-size dependent
$T_C$ from the maximum of the specific heat.
Then we recorded the time series of the energy for each
system size and the associated $T_C$. Fig.~\ref{fig:histo}
shows the resulting energy distribution for $J_2 = 0.3J_1$
and $J_2 = 0.6J_1$. For $J_2 = 0.3J_1$ 
we find a single peak\footnote{The histogram shown
in Fig.~\ref{fig:histo}(a) is very close to a single Gaussian curve, 
as is known for second order phase transitions \cite{P:berg91}.}, consistent
with a conventional second order phase transition.
By contrast, for $J_2 = 0.6 J_1$ a double peak structure
emerges for sufficiently large system size, see Fig.~\ref{fig:histo}(b).
Such a double peak structure is characteristic for a
first order transition \cite{P:challa86,P:berg91,P:borgs92}. The fact that
the double peak structure emerges only for 
large system sizes shows that the transition for $J_2 = 0.6 J_1$ is a weak
first order one. Since we have observed similar behavior for nearby
ratios of $J_2/J_1$, we believe the transition
for $J_2 > J_1/2$ to be of first order,
at least for not too large values of $J_2$.

\begin{figure} 
\begin{center}
\resizebox{\columnwidth}{!}{%
  \includegraphics[]{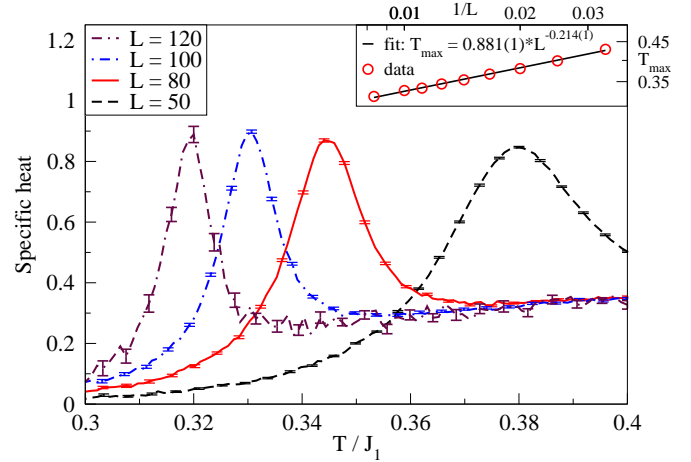}
}
\caption{
Specific heat at $J_2 = 0.5 J_1$ for different lattice sizes
with errorbars at every fourth data point. The peak moves towards
lower temperature (here $T_C/J_1 = 0.380(4), 0.345(4), 0.331(3), 0.320(2)$)
for growing system sizes. The inset shows
$T_C$ (obtained from a polynomial fitting of the maximum
in the specific heat) versus the inverse lattice length in a
\textit{log-log}-scale (errorbars are smaller than the symbols).
The behaviour of $T_C$ is consistent with a power law in $L$,
indicating $T_C=0$ in the thermodynamic limit.}
\label{fig:heat05}
\end{center}
\end{figure}

Fig.~\ref{fig:heat05} shows the specific heat exactly at the critical point
$J_2 = J_1/2$ for different system sizes. The purpose of this computation is
to verify if a direct phase transition from N\'eel to collinear order is possible
at finite temperature, or if any other finite-temperature phase could
exist in this region. Previous publications \cite{P:landau80,P:lopez93}
have shown curves for the specific heat at $J_2 = J_1/2$
with a rounded peak, but we are not aware of any systematic
finite-size analysis. In our results, we observe that
the peaks of the specific heat are moving to lower temperatures for increasing
system sizes, suggesting that the transition temperature is suppressed to $T_C=0$
in the thermodynamic limit. To further substantiate this conclusion, the
inset of Fig.~\ref{fig:heat05} shows
the peak positions ($T_C$) with respect to the inverse lattice length $1/L$.
For the given lattice sizes we find a power law
behaviour for $T_C$ and $1/L$. This is consistent with $T_C = 0$ in the
thermodynamic limit.

\section{Discussion}
\label{sec:discussion}
In this paper, we combined single-spin flip, parallel tempering and
line flip MC algorithms to enhance the efficiency of MC simulations
of the frustrated square-lattice Ising model. These improvements
enabled us to compute critical temperatures in the direct vicinity
of the critical point $J_2 = J_1/2$ where the groundstate is highly degenerate.
For $J_2 < J_1/2$ there is a finite-temperature phase transition
into a N\'eel-ordered state. This transition
belongs to the two-dimensional Ising universality class.
We believe that at the critical point $J_2 = J_1/2$ the transition temperature
is suppressed to zero.
On the right hand side ($J_2 > J_1/2$) there is again a finite-temperature
phase transition into a phase with collinear order.

At $J_2 = 0.6 J_1$ we have observed a double peak structure in the
histogram of the energy. This identifies the phase transition
as a weakly first order one. We believe this first order transition
to be generic for moderate $J_2 > J_1/2$. However, for larger
$J_2$ it becomes increasingly difficult to identify a double peak
structure in the histogram of the energy. Therefore, the transition
could in fact become a second order one for $J_2 \gtrsim J_1$.
Note that a recent MC investigation \cite{P:MKT06} has concluded that the 
transition is second order for $J_2 = J_1$, although in \cite{P:MKT06} only 
smaller lattices have been considered than in the present work.

In order to understand the possible nature of the transition at
large $J_2$, it is instructive to consider the limit $J_1 = 0$
where one has two decoupled Ising models. The critical theory then
consists of two conformal field theories with central charge $c=1/2$ each,
\textit{i.e.}, a $c=1$ theory. Thus, the point $J_1 = 0$
lies within the manifold of $c=1$ conformal field theories where
continuously varying critical exponents are possible in
principle \cite{P:ginsparg88}. However, the coupling given by $J_1$
turns out to be a relevant perturbation of the critical point
such that this coupling should either lead to a first order transition or
a fixed point with $c<1$ \cite{B:cardy96}. According to the
classification of minimal conformal field theories with
$c<1$ \cite{P:BPZ84} a possible second order phase transition
at $J_2 \gg J_1$ should have universal exponents.
Therefore, a scenario with non-universal critical behaviour
\cite{P:SK79,P:landaubinder85,P:MKT06,P:monroe07} does not appear
very plausible from a conformal field theory perspective either.
Our MC results show that large crossover scales exist
in the present model such that very big lattices would be needed for
a reliable numerical determination of the nature of the transition
for $J_2 \gtrsim J_1$.

A similar phase diagram as in the Ising model is also found for the classical $J_1$-$J_2$ 
$X$-$Y$ \cite{P:simon00} and Heisenberg models \cite{P:weber03}. However, we believe that 
the nature of the phase transition for $J_2 > J_1/2$ is a different issue in these two models
due to the different symmetries in spin space.

After having solved the freezing problems in the Ising model,
we are now about to introduce hopping terms and to study
finite-temperature properties of hard-core bosons on the
square lattice using parallel tempering QMC simulations \cite{P:melko07}.

\begin{acknowledgement}
We acknowledge financial support by the Deutsche Forschungsgemeinschaft
under grant No.\ HO~2325/4-1 and through SFB602.
\end{acknowledgement}


\end{document}